\documentclass[a4paper,11pt]{article}

\usepackage{color}

\usepackage{epsfig}

\usepackage{amsfonts}
\usepackage{amsmath}
\usepackage{amssymb}

\newcommand{\be}{\begin{equation}}
\newcommand{\ee}{\end{equation}}

\newcommand{\ba}{\begin{eqnarray}}
\newcommand{\ea}{\end{eqnarray}}

\long\def\symbolfootnote[#1]#2{\begingroup%
\def\thefootnote{\fnsymbol{footnote}}\footnote[#1]{#2}\endgroup}

\begin{document}

\begin{center}
\hfill UB-ECM-PF-07/25

\vspace{0.7cm}

{\Large \bf Secular effects on inflation from one-loop quantum gravity}
\vskip4mm

J.A.~Cabrer\symbolfootnote[1]{jacabrer@ecm.ub.es} and  D.~Espriu\symbolfootnote[2]{espriu@ecm.ub.es}\\
{\it  Departament d'Estructura i Constituents de la Mat\`eria and
 Institut de Ci\`encies del Cosmos,\\ Universitat de Barcelona,\\
 Diagonal 647, 08028 Barcelona, Catalonia, Spain\\
}

\end{center}

\begin{abstract}
In this paper we revisit and extend a previous analysis where the
possible relevance of quantum gravity effects in a cosmological setup was 
studied. The object of interest are non-local (logarithmic) terms generated
in the effective action of gravity due to the exchange in loops of massless
modes (such as photons or the gravitons themselves). We correct one  
mistake existing in the previous work and discuss the issue in 
a more general setting in different cosmological scenarios. 
We obtain the one-loop quantum-corrected evolution equations 
for the cosmological scale factor up to a given order
in a derivative expansion in two particular cases: a matter  
dominated universe with vanishing cosmological constant, and in a de Sitter 
universe. We show that the quantum corrections, 
albeit tiny, may have a secular effect that eventually 
modifies the expansion rate. For a de Sitter universe
they tend to slow down the rate of the expansion, while the effect seems 
to have the opposite sign in a matter dominated universe. 
To partly understand these effects we provide
a complementary newtonian analysis.

\end{abstract}

\section{Introduction}
It has been said \cite{Donoghue} that the effective action of 
quantum gravity is the 
most effective of all effective actions, meaning that an expansion in powers of
$p^2/16\pi M_{P}^2$ would give in normal conditions such a 
tiny contribution to 
any scattering amplitude that the ${\cal O}(p^2)$  (the usual Einstein-Hilbert)
term is good enough for all practical purposes 
(and even for many non-practical ones). 
This is unlike pion physics where the presence of higher order
operators leads lo measurable effects already at moderate energies.
Thus, the fact that the ${\cal O}(p^4)$
terms are ultraviolet divergent does not really bother us in practical
calculations\footnote{In pion physics, and in quantum gravity too, 
one can make sense
of non-renormalizable interactions at the expense of introducing more 
subtraction constants, one per independent operator in the 
derivative expansion.}, although of course the issue is very relevant from
a fundamental point of view.

It is easy to see why quantum corrections are so small. In fact, as already
implied above, the expansion is on powers of  $p^2/16\pi M_{P}^2$ 
(actually 
$\nabla^2/16\pi M_{P}^2$, ${\mathcal R}/16\pi M_{P}^2$) and therefore
very small for physical values of the energy or curvature. Non local
pieces in the effective action ($\sim \ln \nabla^2$), due to the 
presence of strictly massless
modes, somewhat increase the relevance of higher order terms, but locally
they are still negligible. 

There are two reasons why  such apparently hopelessly small
corrections might nevertheless 
be relevant in a cosmological setting. One reason is that
curvature was much larger at early stages of the universe. 
For instance, in 
a de Sitter universe ${\mathcal R}\sim H^2$. In a inflationary
scenario $ H^2= 8\pi G V_0/3$, $V_0$ being the scale of inflation that
is limited by CMB measurements to be $V_0^{1/4} \le 10^{16}$ GeV. 
The Hubble constant $H$  could have been as large as 
$H\sim 10^{13}$ GeV (the present value is $10^{-42}$ GeV). However, even in 
the most favourable case, the correction is still nominally of 
${\cal O}(10^{-12})$ or less and it seems too small to be seen. Or maybe not?

Indeed a second reason to study this problem
is that this nominal suppression overlooks 
the fact that the logarithmic
non local term corresponds to an interaction between geometries 
that is long-range in time, an effect that does not have an 
easy classical interpretation.
 When integrated over time, this may bring about a large enhancement
of the contribution of higher order contributions to the point 
where new interesting affects may appear.

It is quite important to realize that there is no ambiguity in the 
overall coefficient of the
logarithmic non-local term as it depends only on the structure 
of the Einstein-Hilbert
Lagrangian and the number of massless modes (or modes whose mass is
much smaller than the inverse of the horizon radius). Thus the effects are 
model independent and, 
if observable, can be unambigously predicted, at least inasmuch as one can make
accurate predictions in an effective theory.

The possibility of observable effects of the non-local terms in
the effective action in such a situation
was recognized in \cite{EMV}. 
Although the results in \cite{EMV} appear to be correct in substance,
an error slipped in the calculation, unfortunately. This error is corrected
here. More importantly the analysis is extended to different cosmological
models. It is seen that a secular effect from
the non-local terms can be 
predicted unambigously (within the approximations inherent to an effective
action, that is up to terms with higher derivatives) and it is seen to
lead to potentially visible effects.

The relevance of quantum gravity corrections on inflation was also
pointed out in \cite{TW}. Unfortunately, it is difficult to draw
a parallelism between the two approaches. For one thing, we are finding
here a one-loop effect, while the one discussed
in \cite{TW} is a two-loop one due to particle creation \cite{WOO} 
and thus clearly subleading. 

The importance of non-local terms\footnote{By non-local we mean terms 
non-analytic in $\nabla^2$, such as the $\ln \nabla^2$ pieces that appear
in one-loop effective gravity. We do not consider the so-called 
$f({\mathcal R})$ gravity.} in the effective action of gravity cannot
be overemphasized. This has been recently reviewed in \cite{DW}, although 
the non-localities discussed in that paper do not actually correspond to
the one present here. 

In section  \ref{eom} 
we rederive the loop-corrected evolution equation
for the cosmological scale factor in the presence of non-local 
logarithmic terms. In 
section \ref{matter} we apply  
these techniques to a matter dominated universe, governed
by a power-law expansion. In section \ref{cosmoconstant} we rederive
the quantum corrections to the cosmological evolution equation in
a de Sitter background. 
In section \ref{numeric} 
we present the numerical analysis of the solutions and comment on their 
physical relevance. In order to better understand  the 
relevance of the logarithmic terms we
have provided a classical analogy that is presented in section 
\ref{newton}.

The metric convention we use in this paper for Minkowski space is 
\begin{equation}
 \eta_{\mu\nu} = \hbox{diag}(1,-1,-1,-1) .
\end{equation}
The  
Einstein equations are
\begin{equation}
 R_{\mu\nu} - \frac{1}{2} R g_{\mu\nu}= -8\pi G T_{\mu\nu}  
- \Lambda g_{\mu\nu}
,
\label{Einsteinequations}
\end{equation}
where $g_{\mu\nu}$ is the metric tensor, $\Lambda$ is the cosmological 
constant, $R = g^{\mu\nu} R_{\mu\nu}$ and
\begin{gather}
 R_{\mu\nu} = 
\partial_\nu \Gamma^{\alpha}_{\mu \alpha} 
- 
\partial_{\alpha} \Gamma^{\alpha}_{\mu \nu} 
+ 
\Gamma^{\alpha}_{\beta \nu} \Gamma^{\beta}_{\mu\alpha}
- 
\Gamma^{\alpha}_{\beta \alpha} \Gamma^{\beta}_{\mu \nu}
\\
\Gamma^{ \gamma }_{ \alpha \beta } = \frac{1}{2} g^{\gamma \rho} \left( \partial_{\beta }g_{\rho \alpha } + \partial_{ \alpha } g_{\rho \beta } - \partial_{\rho} g_{ \alpha \beta } \right).
\end{gather}
The previous equations are derived from the action 
\begin{equation}
S=\frac{1}{16\pi G} \int dx \sqrt{-g} ({\mathcal R} - 2\Lambda)
+ S_{matter}.
\end{equation}

Quantum corrections to the Einstein-Hilbert action were originally computed by
't Hooft and Veltman in \cite{TV} in the case of vanishing 
cosmological constant,
and by Chistensen and Duff for a de Sitter background \cite{CD}. Other
related
references that we have found particularly useful in the present context are 
\cite{other1} and \cite{other2}.

The key ingredient we shall need is the divergent part of the one-loop
effective action. Using dimensional regularization and setting
$d=4 + 2\epsilon$ we get \cite{other2}
\begin{equation}
\Gamma^{div}_{eff}=
-\frac{1}{16\pi^2\epsilon }\int dx \sqrt{-g}
[c_1{\mathcal R}_{\mu\nu} {\mathcal R}^{\mu\nu}
+c_2\Lambda^2+ c_3 {\mathcal R}\Lambda + c_4{\mathcal R}^2].\label{total}
\end{equation}
The constants $c_i$ are actually gauge dependent and only
a combination of them is gauge invariant.
Using the equations of motion (in absence of matter)
 ${\mathcal R}_{\mu\nu}= g_{\mu\nu}\Lambda$, the previous equation
reduces to the (gauge-invariant) on-shell expression
\begin{equation}
\Gamma^{div}_{eff}=
\frac{1}{16\pi^2\epsilon}\int dx \sqrt{-g}
\frac{29}{5}\Lambda^2.
\end{equation}
If we set $\Lambda=0$ in (\ref{total}), we get the well-known 't Hooft
and Veltman divergence, that in the so-called minimal gauge is
\begin{equation}
\Gamma^{div}_{eff}=
-\frac{1}{16\pi^2\epsilon}\int dx \sqrt{-g}
[\frac{7}{20}{\mathcal R}_{\mu\nu} {\mathcal R}^{\mu\nu}
+ \frac{1}{120}{\mathcal R}^2].\label{tv}
\end{equation}
If the equations of motion are used in the absence of matter this divergence
is absent, as is well known. 

Once the divergence is determined it is straightforward to obtain the
logarithmic pieces since they always appear in the combination
\begin{equation}
\frac{1}{\epsilon} + \ln\frac{\nabla^2}{\mu^2}.
\end{equation}
The dimensionful constant $\mu$ is the substraction scale 
that is required for dimensional consistency.

\section{\label{eom}The equations of motion in the presence of non-local
terms}

In this section we shall derive quantum corrected equations of motion
for the cosmological scale factor including the non-local logarithmic terms
that appear in the one-loop effective action.
We shall consider here a simplified effective action that includes 
only terms containing the scalar curvature 
we split the action into three parts 
and redefine the constants for convenience
\begin{align}
 \notag
 S &= \kappa^2 \left( \int dx \sqrt{-g} \mathcal{R} 
+ \tilde{\alpha} \int dx \sqrt{-g} \mathcal{R} \ln ( \nabla^2 / \mu^2 ) 
\mathcal{R} + \tilde{\beta} \int dx \sqrt{-g} \mathcal{R}^2 \right)
 \\
 &\equiv
 \kappa^2 \left( S_1 + \tilde\alpha S_2 + \tilde\beta S_3 \right),
 \label{effaction}
\end{align} 
where $\kappa^2=M^2_{P}/16\pi=1/16\pi G$. $\mu$ is the subtraction
scale whose contribution is by itself local, but gives 
the right dimensions
to the non-local term.   
The coupling $\tilde\beta$ is $\mu$ dependent in such a way that 
the total action $S$ is $\mu$-independent. While the value
of $\tilde\beta$ is actually dependent on the UV structure of the
theory (it contains information on all the modes -massive or not- that
have been integrated out), the value of $\tilde\alpha$ is unambiguous as 
it depends only on the IR structure of gravity, that is entirely described
by the Einstein-Hilbert Lagrangian
and the massless modes.

In conformal time, $dt = a d\tau$, we have
\begin{equation}
g_{\mu\nu}=a^2(\tau)\eta_{\mu\nu},\ 
\mathcal{R}=6 \frac{{a''(\tau)}}{a^3(\tau)},\
\sqrt{-g}=a^4(\tau).
 \label{conformalg}
\end{equation}

In order to obtain the modified equations of motion for the
cosmological scale factor,
we first perform the variation of the local action, 
getting the following results
\begin{equation}
 \frac{\delta S_1}{\delta a(\tau)} = 12 a''
\end{equation}
\begin{equation}
 \frac{\delta S_3}{\delta a(\tau)} = 
72 \left( -3 \frac{(a'')^2}{a^3} - 4 \frac{a' a'''}{a^3} 
+
 6 \frac{(a')^2 a''}{a^4} + \frac{a^{(4)}}{a^2} \right)
\label{variationS3}
\end{equation}

The d'Alembertian in conformal space is related to the 
Minkowski space operator by
\begin{equation}
\nabla^2={a}^{-3}\Box\, a +\frac{1}{6} \mathcal{R}
\end{equation}
Neglecting the $\mathcal{R}$ term in the previous equation and 
commuting the scale factor $a$ with the flat d'Alembertian (terms
with higher derivatives are neglected in the effective action philosophy), 
we can write
\begin{equation}
\nabla^2=\left(\frac{a}{a_0}\right)^{-2}\Box
\end{equation}
Where $a_0=a(0)$. With this rescaling (absorbable in $\tilde{\beta}$), at $\tau=0$
the d'Alembertian in conformal space matches with the Minkowskian one.
We will set $a_0=1$ for simplicity.

We can now separate $S_2$ in turn into a local and a genuinely non-local piece
\begin{align}
\notag
S_2  &=  \int dx\sqrt{-g}\ \left(-2\mathcal{R}\ln(a)\mathcal{R}+\mathcal{R}\ln(\Box/\mu^2)\mathcal{R}\right)\notag
\\
& \equiv S_2^I+S_2^{II}.
\end{align}
The variation of $S_2^I$ gives
\begin{align}
\frac{\delta S_2^I}{\delta a(\tau)}  
=&
-72\left\lbrace 
\frac{(a')^2 a''}{a^4}  \left[ 12 \ln(a) - 10 \right]
+ 
\frac{a' a'''}{a^3} \left[ -8 \ln(a) + 4 \right]
+
\right.
\\
&\left.
+
\frac{(a'')^2}{a^3} \left[ -6 \ln(a) + 2 \right]
+
\frac{a^{(4)}}{a^2} 2 \ln(a)
\right\rbrace
\end{align}
In order to determine the variation of $S_2^{II}$ we 
need to compute the Green function 
\begin{equation}
\langle x\vert \ln \Box\vert y\rangle.\label{gf}
\end{equation}
We follow
the method of \cite{EMV} that we shall not repeat here.
We mention here that the normalization of the delta function used
in \cite{EMV} was non-covariant\footnote{We thank G.P\'erez for
bringing this to our attention}. Using the proper
normalization and correcting for this mistake we find 
\begin{equation}
 S_2^{II} = 36 \int d\tau \frac{a''(\tau)}{a(\tau)} 
\int_0^{\tau} d\tau' \frac{1}{\tau-\tau'} \frac{a''(\tau')}{a(\tau')}
\end{equation}

The variation of $S_2^{II}$ is
\begin{align}
\notag
 \frac{\delta S_2^{II}}{\delta a(\tau)} 
=
36 &\left\lbrace 
\left[ 2 a^{-3}(\tau)\left(a'(\tau)\right)^2 
- 2 a^{-2}(\tau) a''(\tau) \right] 
\int_0^\tau d\tau' \frac{1}{\tau - \tau'} \frac{a''(\tau')}{ a(\tau')}
\right.
\\
&\left.
- 2 a^{-2}(\tau) a'(\tau) \frac{\partial}{\partial \tau} 
\left(\int_0^\tau d\tau' \frac{1}{\tau - \tau'} \frac{a''(\tau')}{ a(\tau')}
\right) + 
a^{-1}(\tau) \frac{\partial^2}{\partial \tau^2} 
\left( \int_0^\tau d\tau' \frac{1}{\tau - \tau'} \frac{a''(\tau')}{ a(\tau')} 
\right) 
\right\rbrace
\end{align}
Using repeatedly integration by parts it is possible to further simplify
the expression eliminating the derivatives acting on the integrals
\begin{gather}
\notag
 \frac{\partial}{\partial\tau} \left( \int_0^\tau d\tau' \frac{1}{\tau - \tau'} \frac{a''(\tau')}{ a(\tau')} \right) 
=
\frac{a_0^{-1} a_0^{\prime \prime}}{\tau} 
+
\\
+
\int_0^\tau d\tau' \frac{1}{\tau - \tau'} \left[ - a^{-2}(\tau') a'(\tau') a''(\tau') + a^{-1}(\tau') a'''(\tau') \right]
\end{gather}
\begin{gather}
\notag
 \frac{\partial^2}{\partial\tau^2} \left( \int_0^\tau d\tau' \frac{1}{\tau - \tau'} \frac{a''(\tau')}{ a(\tau')} \right) 
=
-\frac{a_0^{-1} a_0^{\prime \prime}}{\tau^2} 
+
\frac{-a_0^{-2} a_0^{\prime} a_0^{\prime\prime} + a_0^{-1} a_0^{\prime\prime\prime}}{\tau}
+
\\ \notag
+
\int_0^\tau d\tau' \frac{1}{\tau - \tau'} \left[
2 a^{-3}(\tau') (a'(\tau'))^2 a''(\tau') - a^{-2}(\tau') (a''(\tau'))^2 
\right.
\\
\left.
- 2a^{-2}(\tau') a'(\tau') a'''(\tau') 
 + a^{-1}(\tau')a^{(4)}(\tau')
\right]
\end{gather}
The terms that potentially diverge at $\tau=0$ arise from the fact that 
we patch together Minkowski and FRW spaces at $\tau=0$. 
If it is done smoothly enough, the derivatives 
of the scale factor should vanish at that point. 

The previous expressions are formal in the sense that the 
short distance singularities have not been properly regularized. To regulate
the region where $\tau\to\tau^\prime$ we use dimensional regularization as
explained in detail in \cite{EMV}. The regulator is 
in fact introduced when the Green function (\ref{gf}) is computed in
$4+2\epsilon$ dimensions. In
practice, this amounts to the replacement
\be
\frac{1}{\tau-\tau^\prime}\to 
\frac{\mu^{-2\epsilon}}{(\tau-\tau^\prime)^{1+2\epsilon}},
\label{regularization}
\ee
where $\mu$ is the subtraction scale previously introduced. The physical
result is obtained for $\epsilon\to 0$. The logarithm of the effective
action (\ref{effaction}) is then  
reproduced and the $1/\epsilon$ divergence 
(proportional to ${\mathcal R}^2$) 
is cancelled by the (divergent) counterm included in $\tilde\beta$.
Thus it is easy to compute, even numerically, the genuinely non-local
piece, as the would-be singular term in the integral is clearly identified.   

At this point there are several ways to proceed. One might of course
attempt to find solutions of the equation of motion for $a(\tau)$ obtained by
adding the variations for $S_1$, $\tilde\beta S_3$ and
$\tilde\alpha (S_2^I + S_2^{II})$ that we have just computed. 
This way of proceeding is not really justified if the ${\cal O}(p^4)$ terms
are understood as a correction.

In the spirit of effective Lagrangians it is better to proceed otherwise. We
obtain first the lowest order equation of motion from $S_1$ and plug it in 
$\tilde\alpha (S_2^I + S_2^{II}) + \tilde\beta S_3$. The quantum corrections
act then as an external driving force superimposed to Einstein
equations. This procedure of course gives trivially a net zero additional
contribution in the present toy model
as neither matter nor a cosmological constant have been considered.
In the next sections we shall introduce $T_{\mu\nu}$ and $\Lambda$
to find more interesting effects.

\section{\label{matter} Quantum gravity effects in a matter dominated universe}

Let us consider a pressureless distribution of matter characterized
by the energy-momentum tensor $T_{\mu\nu}= {\rm diag}( \rho,0,0,0)$. We can use
Einstein equations to write (in conformal time)
\begin{equation} 
{\mathcal R}_{00}= -4\pi G a^2(\tau) \rho,\qquad
{\mathcal R}_{ij}= 4\pi G a^2(\tau) \rho.
\end{equation}
The addition of an energy-momentum tensor leads in principle to the appearence
of ${\cal R} T$, $T^2$ and $ T_{\mu \nu}T^{\mu\nu}$ terms in the 
effective action. However, the details of the calculation are still controversial
to some extent. Here we shall adopt the view that a classical energy-momentum
tensor (such as one representing presureless dust) interacts only via classical
gravitons. 

We then substitute these expressions in the non-local ${\cal O}(p^4)$ term
effective action
\begin{align}
S&=\frac{1}{16\pi G} \int dx \sqrt{-g} {\mathcal R} \\
&-\frac{1}{16\pi^2}\int dx \sqrt{-g}
\left[\frac{7}{20}{\mathcal R}^{\mu\phantom{\nu}}_{\phantom{\mu}\nu}\ln\frac{\nabla^2}{\mu^2}
 {\mathcal R}^{\nu\phantom{\mu}}_{\phantom{\nu}\mu}
+ \frac{1}{120}{\mathcal R}\ln\frac{\nabla^2}{\mu^2}{\mathcal R}\right]\\
&+\hbox{local terms of } {\cal O}(p^4) + S_{matter}.
\end{align}
The local ${\cal O}(p^4)$ terms (proportional 
to ${\mathcal R}^2$ and ${\mathcal R}^{\mu\phantom{\nu}}_{\phantom{\mu}\nu} {\mathcal R}^{\nu\phantom{\mu}}_{\phantom{\nu}\mu}$ 
can be used to absorb the divergences of the one-loop action. If one 
day one would be able to make a precise measurement of some quantum
gravity effect, one could determine these coefficients and all other
predictions would be calculable unambiguously with an  ${\cal O}(p^4)$
precision. For the time being, we only know unambigously the logarithmic
non-local terms. But these are the ones that are like to give enhanced
contributions as we discussed in the introduction.

After use of the lowest order, ${\cal O}(p^2)$ equations of motion, the
non local terms simplify considerably
\begin{align}
S &= \kappa^2\left( \int dx \sqrt{-g} {\mathcal R}
+\tilde\alpha \int dx \sqrt{-g} \rho \ln\frac{\nabla^2}{\mu^2} \rho\right)
+ S_{matter}\\
&\equiv  \kappa^2\left( \int dx \sqrt{-g} {\mathcal R}
 + \tilde\alpha S_2\right) +S_{matter}.
\end{align}
We omit the purely local part (analogous to $S_3$ in the previous 
section) as is not calculable from the low energy information only 
and logs dominate anyway.

The value of $\tilde\alpha$ can be determined
readily from the results of 't Hooft and Veltmann \cite{TV} (which 
includes only the contribution of virtual gravitons) and
the lowest order
equations of motion in the presence of matter
\begin{equation}
 \tilde\alpha= -16\pi G^3 \times \frac{43}{30}.
\end{equation}
The correction
from massless photons or Yang-Mills fields 
does not seem to change the sign of $\tilde\alpha$
\cite{DTN}. We
have not considered other possibilities.  
In fact, the precise value of $\tilde\alpha$ is not so important (but the
sign and its rough order of magnitude is).

$S_2$ modifies the equations of motion by adding a new term
that is simply
\begin{equation}
\frac{\delta S_2}{\delta a(\tau)}
= -2 \rho(t)^2a^3(\tau) [4\ln (a(\tau)) + 1] 
+ 2\rho(\tau) a(\tau) \int_0^\tau \rho(\tau^\prime) a^2(\tau^\prime)
\frac{\mu^{-2\epsilon}}{\vert \tau -\tau^\prime\vert^{1+2\epsilon}}, 
\end{equation}
multiplied by $\tilde\alpha$.
Notice that quantum effects introduce long range interactions in time
between global matter densities at different times.

The value of $\rho(\tau)$ is known from the lowest order equation
of motion. In conformal time
\begin{equation}
\rho(\tau)\sim a^{-3}(\tau),\qquad a(\tau)\sim \tau^2. 
\end{equation}
We shall discuss the physical relevance of these corrections after
discussing in detail the solution in the de Sitter case.

\section{\label{cosmoconstant} Quantum gravity effects in a de Sitter universe}

We shall proceed in a way similar to the previous section, but we shall
now omit the energy-momentum tensor in the lowest order equations of motion.
In fact it is not correct to simply 
assume $T_{\mu\nu}\sim g_{\mu\nu}\Lambda$ 
in the equations of motion and just use the previous formulae (that would
roughly be equivalent to exchanging $\rho $ by $\Lambda$ in the previous 
section, up to constants). The cosmological constant is the most relevant
operator in gravity and it must be introduced from the outset.

The relevant one-loop corrected effective action is 
\begin{align}
S&=\frac{1}{16\pi G} \int dx \sqrt{-g} ({\mathcal R}-2\Lambda) 
+\frac{1}{16\pi^2}\int dx \sqrt{-g} \frac{29}{5} 
\Lambda \ln\frac{\nabla^2}{\mu^2} 
\Lambda + {\rm local ~terms~of~} {\cal O}(p^4)\\
&\equiv  \kappa^2\left( \int dx \sqrt{-g} ({\mathcal R}-2\Lambda)
 + \tilde\alpha S_2\right).
\end{align}
Now $\tilde\alpha$ is very different
from the previous case
\begin{equation}
 \tilde\alpha= \frac{G}{\pi}\times \frac{29}{5}
\end{equation}
The dimensions of $\tilde\alpha$ are of course different 
as the dimensionality of $\rho$ and
$\Lambda$ is not the same. Most importantly, it has the opposite sign. 
Notice that all our expressions are written in such a way that
it is possible to consider a time-dependent cosmological constant (or
matter density).

We split $S_2$ in two parts
\begin{gather}
 S_2^{I} = -2 \int dx \sqrt{-g} \Lambda^2\ln(a) \\
 S_2^{II} = \int dx \sqrt{-g} \Lambda \ln(\square/\mu^2) \Lambda, 
\end{gather}
and obtain the corresponding variations following the 
method outlined in section \ref{eom}
\begin{equation}
 \frac{\delta S_2^I}{\delta a(\tau)} = -2 \Lambda^2 a^3(\tau) \left[4\ln(a(\tau)) + 1\right]
\end{equation}
\begin{equation}
 \frac{\delta S_2^{II}}{\delta a(\tau)} =  2 \Lambda^2 a(\tau) \int_0^\tau d\tau' a^2(\tau') \frac{\mu^{-2\epsilon}}{\vert \tau - \tau' \vert^{1+2\epsilon}}.
\end{equation}

The equation of motion will be 
\begin{equation}
12 a^{\prime\prime}(\tau) - 8 \Lambda a^3(\tau) 
+\tilde\alpha 
\frac{\delta S_2}{\delta a(\tau)}=0, \label{eqcc}
\end{equation} 
which at lowest order is just
\begin{equation}
12 a^{\prime\prime}(\tau) - 24 H^2 a^3(\tau)= 0,\label{loweqmotion}  
\end{equation}
where $H^2=\Lambda/3$.
The solution of (\ref{loweqmotion}) is
\begin{equation}
a_I(\tau)=\frac{1}{ 1-H \tau}.
\label{scalefactorconformal}
\end{equation}
The final step to solve 
iteratively (\ref{eqcc}) 
is to plug the $0$-th order solution $a_I(\tau)$ into the 
variation of $S_2$ and recalculate the solution for $a(\tau)$. 

It is clear that, apart from the sign difference, the quantum
effects are formally very similar for a matter dominated and for a de Sitter
universe. However, the fact that the signs are opposite means that their
back reaction is completely opposite. If quantum corrections enhance
expansion in one case, they will slow it down in the other. Futhermore,
the size of the correctios is very different: the corrections in a matter
dominated universe are down by a factor $H^2/M_P^2$ with respect to the
ones in a de Sitter space-time with a large cosmological constant, already
expected to be small. Let
us now investigate the numerical relevance of the latter ones.

\section{\label{numeric} Solving the evolution equation}

As we just discussed, 
we proceed by solving the varied gravitational action by a
perturbative approximation, i.e., we consider the non-standard terms
as a correction to the standard inflationary solution. This 
perturbative procedure
is of course only valid as long as 
the correction is small compared to the unperturbed
solutions.

Before doing that we find it convenient to change time coordinates by
introducing a variable $s$ defined through $a_I = e^s$.  Then
\begin{equation}
 \left. \frac{\delta S_2^I}{\delta a(\tau)} \right\vert_{a_I} = -2 \Lambda^2 e^{3s} \left[4s + 1\right]
\end{equation}
\begin{equation}
 \left.\frac{\delta S_2^{II}}{\delta a(\tau)}\right\vert_{a_I}
  = 
  2 \Lambda^2 e^s I(s)
\end{equation}
and the equation of motion reads
\begin{equation}
 e^{2s} a''(s) + e^{2s} a'(s) - 2 a^3(s) = \frac{3}{2} \tilde{\alpha} H^2 \left( - e^{3s} (1 + 4s) + e^{s} I(s) \right),
\end{equation}
where $I$ is defined in conformal time as
\begin{equation}
I(\tau) \equiv \mu^{-2\epsilon} \int_0^\tau d\tau' \frac{a_I^2 (\tau')}{(\tau 
- \tau')^{1 + 2\epsilon}} = 
- \frac{1}{2\epsilon} (\tau \mu)^{-2\epsilon} \,_2F_1 (1 , 2 ; 1 - 2\epsilon ; 
H\tau) 
\end{equation}
with $\,_2F_1$ being a hypergeometric function. 
Let us expand this expression around $\epsilon=0$ to the first order, 
disregarding higher orders since we eventually
 take the limit $\epsilon\rightarrow0$. 
Using $\,_2 F_1 ( 1 , n ; 1 ; H\tau) = (1 - H\tau)^{-n} = a_I^n(\tau)$, we get
\begin{align}
\notag
 I(\tau)
&= \ln(\tau\mu) \,_2 F_1 ( 1 , 2 ; 1 ; H\tau) - 
\frac{1}{2} \left.\frac{\partial}{\partial \epsilon} 
\Big[ \,_2F_1 ( 1 , 2 ; 1 - 2\epsilon ; H \tau) \Big]
\right\vert_{\epsilon=0}
=
\\
&=
\ln(\tau \mu) a_I^{2}(\tau) - 
\frac{1}{2} \left.\frac{\partial}{\partial \epsilon} 
\Big[ \,_2F_1 ( 1 , 2 ; 1 - 2\epsilon ; H \tau) \Big]
\right\vert_{\epsilon=0}.
\end{align}
This can be computed and written in $s$ time as
\begin{equation}
 I(s) =  \ln\left(\frac{\mu}{H} (1 - e^{-s})\right) e^{2s} + e^s(1 - e^s - se^s),
\end{equation}
and the equation to solve is 
\begin{equation}
  e^{2s} a''(s) + e^{2s} a'(s) - 2 a^3(s) = \frac{3}{2} \tilde{\alpha} H^2 \left[ - (5 s + 2) e^{3s} + e^{2s} +
   e^{3s} \ln\left(\frac{\mu}{H} (1 - e^{-s}) \right) \right]
   \label{eqnmotion2}
\end{equation}
Note that $\tilde\alpha$ appears only in the combination
$\tilde\alpha H^2$. Since there are $H$ large uncertainties
in $H$ in practice only the sign of $\tilde\alpha$ is relevant. In addition,
there is some ambiguity associated to the choice of the renormalization
scale that appears in the combination $\ln(\mu/ H)$ \footnote{Recall
that a change of $\mu$ is equivalent to a redefinition of the
local ${\cal O}(p^4)$ counterterms}. The dependence on the 
subtraction scale is mild (logarithmic) but it is inherent to 
the effective action approach. To reverse the sign of the effect
one has to go to absolutely unreasonable values of $\mu$.

\begin{figure}[t]
\centering
 \includegraphics{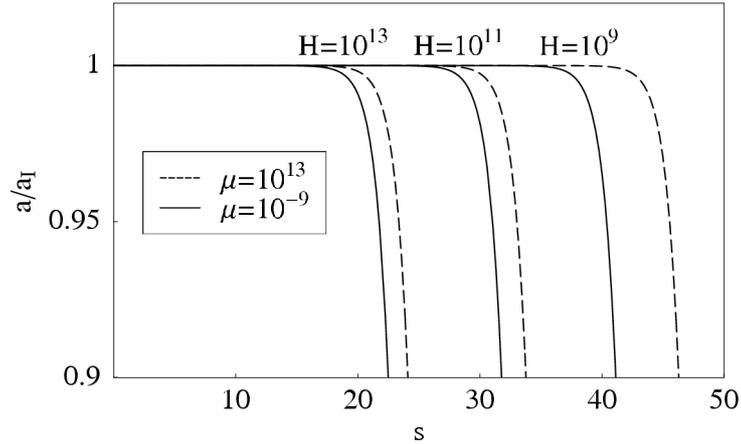}
 \caption{The scale factor relative to the inflationary expansion 
for different values of the renormalization scale $\mu$ and the Hubble constant $H$ (all units are $\hbox{GeV}$). $s$ gives roughly the number of e-folds. The precise definition is given in the text.}  
 \label{fig:graph1}
\end{figure}

Eq. (\ref{eqnmotion2}) can now be solved for different 
values of $H$ and $\mu$, where $H \lesssim 10^{13}\,\hbox{GeV}$. 
The solution is shown in Fig. \ref{fig:graph1}. We can see that 
the curves present a very similar behaviour for the different values shown, 
though a higher value of $H$ leads earlier to deviations from the usual inflationary expansion. Higher values of $\mu$ also have this effect, which is larger as $H$ increases (note that what is relevant in the equation is the ratio $\mu/H$). In fact, we can see from (\ref{eqnmotion2}) that if we considered values of $\mu/H$ large enough (but not relevant physically), the logarithm term would become dominant and the deviation would be positive. 

\section{\label{newton}Newtonian approximation}

It is well known that one can derive the 
Friedmann equations using only newtonian physics \cite{Lyth}. 
We will redo this exercise considering quantum corrections to the Newton 
potential. These have been computed by several authors \cite{corrections}.
In fact, the magnitude of this correction and even the sign has been 
the subject of a long controversy. The correct value for the quantum
correction to the gravitational force between two masses is given
for instance in \cite{AS2006}. This we believe to be the relevant 
modification in the context of this newtonian approximation, but we
shall redo the analysis for an arbitrary value of the correction for
the sake of completeness.
 
Quantum corrections can be summarized
in a modification of  Newton's `constant' in the following way
\begin{equation}
 G(r) = G \left( 1 + \xi \frac{G\hbar}{r^2} \right),
\end{equation}
where $\xi$ is a dimensionless constant.

\begin{figure}[t]
\centering
 \includegraphics{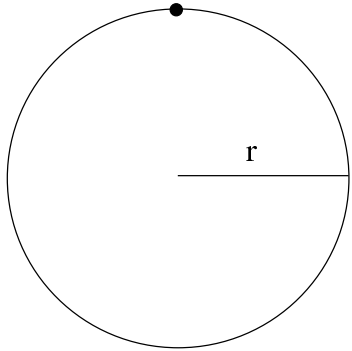}
 \caption{In order to derive the Friedmann equation, we consider a 
test particle on the surface of a virtual sphere in our infinite, 
homogeneous and isotropic universe.}
 \label{fig:sphere}
\end{figure}

We consider an infinite, homogeneous and isotropic expanding universe. 
We will describe the equations of movement for a test particle on the 
surface of a comoving sphere, expanding with the universe. 
The mass inside the sphere is $\rho (4\pi/3) r^3$, where $\rho$ is 
the density of the universe and $r$ the radius of the sphere. 
(see Fig. \ref{fig:sphere}). 
The kinetic energy of the (unit mass) particle is $(1/2)\dot{r}^2$. 
In order to compute its potential energy $U$ we shall consider first the 
force between our test particle and another particle with mass $m$ 
at a distance $l$ from it
\begin{equation}
 \vec F = - G(r)m\frac{\hat l }{l^2} = - G m \frac{\hat l}{l^2} 
- \xi  G^2\hbar m \frac{\hat l}{l^4} \equiv \vec F_1 + \vec F_2
 \label{eq:energydistl}
\end{equation}
The first part $\vec F_1$ is of course elementary and once integrated 
it leads to a contribution to the total potential energy $U=U_1+U_2$ 
that is easy to guess  
\begin{equation}
 U_1 = - \frac{G M }{r} = - \frac{4\pi G \rho}{3} r^2
\end{equation}
For the second part, we have to integrate $\vec F_2$ over the whole 
sphere and also the space outside of it, since we can not make use of Gauss 
theorem. After doing this we get the following result 
for the contributions from matter outside and inside the sphere, respectively
\begin{align}
 \notag
 \vec{F}_{2_{out}} &=  \frac{4\pi}{3} \left(\xi  G^2\hbar \right)  \rho \left( \left. 
\frac{r}{r^2 - \tilde{r}^2} \right\vert_{\tilde{r} \rightarrow r^+}\right) 
\hat{r}
\\
 \vec{F}_{2_{in}} &= - \frac{4\pi}{3}  \left(\xi  G^2\hbar \right) \rho 
\left( \frac{1}{r} +
\left. \frac{\tilde{r}}{\tilde{r}^2 - r^2} \right
\vert_{\tilde{r} \rightarrow r^-}
\right) 
\hat{r}
\end{align}
Where $\hat{r}$ is the unit radial vector pointing to increasing values of $r$. If we sum these two contributions, the divergences in the boundary cancel 
and we have
\begin{align}
 \notag
 \vec{F}_2 &= - \left(\xi  G^2\hbar \right) \frac{M}{r^4} \hat{r}
 \\
 U_2 &=  - \left(\xi  G^2\hbar \right) \frac{M}{3r^3} =  
- \frac{4\pi}{3} \left(\xi  G^2\hbar \right) \frac{\rho}{3}
\end{align}
This is a manifestation of Birkhoff's theorem.

We can now write the total energy of our test particle, that is
\begin{equation}
E = \frac{1}{2} \dot{r}^2 - \frac{4\pi G \rho}{3} r^2 
- \frac{4\pi}{3} \left(\xi  G^2\hbar \right) \frac{\rho}{3}
\end{equation}
Writing $r(t)=a(t)x$ and dividing by $a^2/2$ we get the modified 
Friedmann equation.
\begin{equation}
 \left(\frac{\dot{a}(t)}{a(t)}\right)^2 = \frac{8\pi G}{3} \rho 
+ \frac{8\pi}{3} \left(\xi  G^2\hbar \right) 
\frac{\rho}{3} \frac{1}{a^2(t) x^2} - \frac{K}{a^2(t)}
\end{equation}
Where $K = -2E/x^2$ is the curvature, which we set to $0$ since 
we are considering a flat space. As we see, the local coordinate $x$ appears
explicitly and in order to put precise numbers we have to be very precise as we 
define our unit system. This is not very relevant for practical purposes due
to the smallness of the correction and we shall embed this into $\xi$.

After identifying the cosmological constant $\Lambda = 8\pi G \rho$, we have
\begin{equation}
 \left( \frac{\dot a(t)}{a(t)} \right)^2 = \frac{\Lambda}{3}
\left( 1 + \frac{\xi G \hbar}{3} \frac{1}{a^2(t)}\right).
\end{equation}

In order to compare with the time coordinate we have introduced before,
we revert to the variable $s$, remembering that, to the lowest order, $H^2 = \Lambda/3$. 
\begin{equation}
 (a'(s))^2 e^{2s} = a^4 (s) \left( 1 + \frac{\xi G \hbar}{3} \frac{1}{a^2(s)} \right)
\end{equation}

\begin{figure}[t]
\centering
 \includegraphics{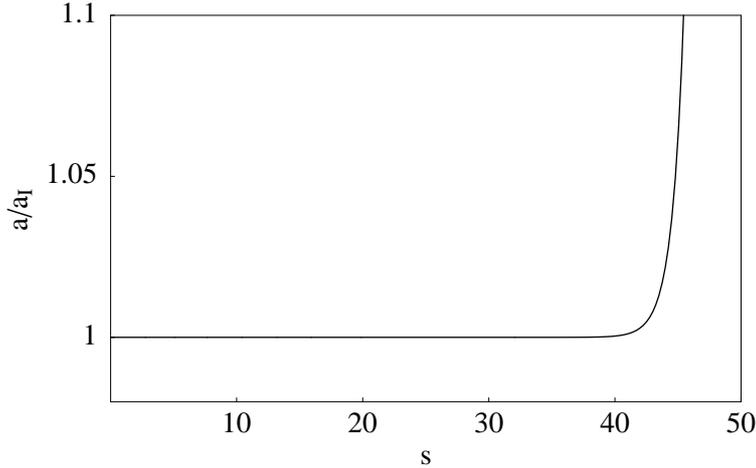}
 \caption{The scale factor relative to the inflationary expansion resulting from our Newtonian approach. This curve corresponds to $\xi G \hbar\sim 10^{-20}$.}
 \label{fig:graph2}
\end{figure}

The resulting evolution can be computed easily and the effects are seen to
be very small; so small in fact that to make them visible in a reasonable time evolution (comparable to the times seen in Fig. \ref{fig:graph1})  we have taken
$\xi G \hbar\sim 10^{-20}$, which is an unreasonable value for sure (see Fig. \ref{fig:graph2}). The
effect is to increase the expansion rate. This simple exercise teaches
us two things. One is that it is incorrect to simple take
the cosmological constant as part of the `matter balance' of the universe
and stick to newtonian physics. Incidentally, the
calculation would be the same for a matter dominated universe and the effects
even tinier. The second thing we learn is that classical physics with
a potential modified by quantum corrections
is unable to reproduce the enhancement that the non-local terms bring
about due to long range time correlations.

\section{Conclusions}

In this paper we have analyzed the relevance of the non-local
quantum corrections due to the virtual exchange of gravitons and other
massless modes to the 
evolution of the cosmological scale factor in FRW universes. We have
considered two different setups: a matter dominated universe, characterized
by a matter density $\rho(t)$, and a de Sitter universe with a large
cosmological constant.

In the de Sitter universe,
while the effects are locally absolutely tiny, even after allowing for
the largest possible value of $H$, we have found that they lead to a
noticeable secular effect that slows down the inflationay expansion
after a long time. This is a pure one-loop effect that is not actually
related to particle creation and its back-reaction on the universe
expansion and which constitutes a two loop effect as emphasized
by Tsamis and Woodard. In a matter dominated universe the effect
is a lot smaller, and it may be of the opposite sign. It is 
quite interesting
that quantum effects seem to enhance the expansion rate in this case.
 
The physical effects thus depend crucially on the sign of the quantum 
corrections. We have also seen that this effect has no really classical
analogy.

It is quite important to emphasize once more that the results presented
here are not `just another model'. Quantum gravity non-local 
loop corrections are 
required by unitarity, even if the theory is non-renormalizable;
 they can be computed quite precisely in a derivative expansion; 
and they appear to be of some relevance in the present
situation. Perhaps
the most interesting result of our analysis is indeed the fact that these
effects can be predicted unambiguously within the limits of an effective
theory.

\section*{Acknowledgements}

We would like to thank G. P\'erez for pointing out to us the mistake
in the original calculation and for several discussions concerning 
the possible relevance of non-local terms. We acknowledge also 
discussions with Y. Khriplovich, T. Multamaki, E. Vagenas and E. Verdaguer
as well as an early conversation with J. Donoghue. 
The financial support
received through grant FPA2004-04582  and the European RTN grant ENRAGE is gratefully
acknowledged. The work of J.A. Cabrer has been supported by a grant from
the Spanish Ministry of Education and Science.
D.E. would like to thank the hospitality of the CERN PH Division where
this work was completed.

\end{document}